# Performance Evaluation of Vanilla, Residual, and Dense 2D U-Net Architectures for Skull Stripping of Augmented 3D T1-weighted MRI Head Scans

Anway S. Pimpalkar[1][0000-0003-2535-2697], Rashmika K. Patole[1][0000-0001-7247-0296], Ketaki D. Kamble[1][0000-0003-4674-0488] and Mahesh H. Shindikar[1][0000-0001-8845-5488]

[1] COEP Technological University, Pune MH 411005, India
pimpalkaras19.extc@coep.ac.in, rkp.extc@coep.ac.in,
kamblek.appsci@coep.ac.in, smh.appsci@coep.ac.in

**Abstract.** Skull Stripping is a requisite preliminary step in most diagnostic neuroimaging applications. Manual Skull Stripping methods define the gold standard for the domain but are time-consuming and challenging to integrate into processing pipelines with a high number of data samples. Automated methods are an active area of research for head MRI segmentation, especially deep learning methods such as U-Net architecture implementations. This study compares Vanilla, Residual, and Dense 2D U-Net architectures for Skull Stripping. The Dense 2D U-Net architecture outperforms the Vanilla and Residual counterparts by achieving an accuracy of 99.75% on a test dataset. It is observed that dense interconnections in a U-Net encourage feature reuse across layers of the architecture and allow for shallower models with the strengths of a deeper network.

## 1 Introduction

Neuroimaging is a prevalent field for diagnostic assessments of neuroanatomy, neurophysiology, and internal functions such as cognition and control. Numerous techniques are employed in neuroimaging, such as Computed Tomography (CT), Magnetic Resonance Spectroscopy, Magnetization Transfer Imaging, Cerebral Perfusion Imaging, Single Photon Emission Computed Tomography (SPECT), Positron Emission Tomography (PET), Magnetic Resonance Imaging (MRI), Diffusion Tensor Imaging (DTI), and Ultrasound Imaging [1]. MRI scanning is used extensively in medical diagnostic settings due to its noninvasive and nondestructive nature. It employs static and variable external magnetic fields to perturb hydrogen atoms in the brain tissue. The alignment of the magnetic field of the hydrogen atoms changes to and from its original position once the external field is cut out or induced, causing the emission of a signal that receivers can pick up. The intensities of the received signals represent the different tissues present in the scan [2].

Different image types are created by varying the sequence of the radio pulse frequency applied. Therefore, MRI scans can be imaged under different contrasts, namely (1) T1 weighted; (2) T2 weighted; and (3) Fluid Attenuated Inversion Recovery (FLAIR). T1-weighted MRIs rely upon the longitudinal relaxation of the magnetic vectors of a tissue after a pulse frequency is cut-off. The time taken for the vectors to return to the original vector direction is different for the different tissues present in the brain, leading to a variation in the signal intensities emitted [3]. T2-weighted MRIs measure the transverse relaxation of the magnetic vectors when a pulse frequency is induced, the different rates of which correspond to the different neural tissues [4]. FLAIR MRI scans are similar to T2-weighted techniques, with the additional removal of Cerebrospinal Fluid (CSF) from the resulting images. This sequence makes the differentiation between CSF and abnormalities much easier [5].

A structural MRI images the entirety of the tissue present in the head. Most neuroscience studies involving the brain require a segmentation step, which extracts the brain tissue from the scan, often referred to as Skull Stripping. It involves removing non-cerebral tissue from a head scan. It may subtract any non-cerebral tissue from the scan, including the skull, meninges, and scalp [6]. It is a critical step in most pre-processing workflows, as rectifications in the segmentation map cannot be performed in subsequent steps.

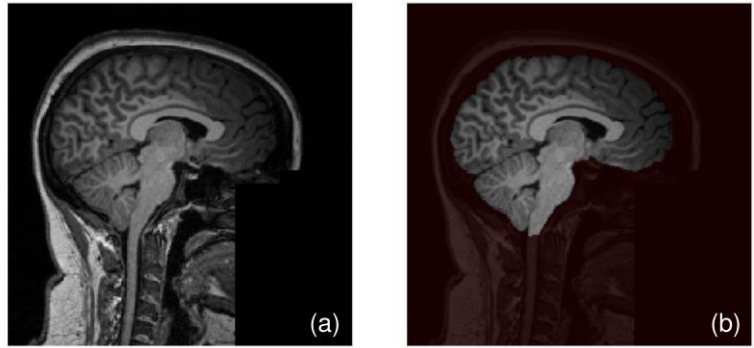

**Fig. 1.** A representation of (a) Sagittal 3D T1-w MRI slice from the NFBS repository [7]; and (b) corresponding Skull Stripped mask superimposed on the MRI scan.

Manual segmentation held prominence in the initial days of Skull Stripping. However, manual delineation methods are time-consuming and cannot be used for developing pipelines that involve multiple pre-processing steps. Thus, research into semi-automated and automated methods began to rise, and since, these methods have taken prevalence in the field. However, manual segmentation is still employed to define the gold standard or ground truth for comparison metrics to the advanced methods used today [8].

Conventional methodologies used for Skull Stripping are classified into five types: (1) morphology-based; (2) deformable surface-based; (3) intensity-based; (4) template and atlas-based; and (5) hybrid. Modern methods can be classified into (1) machine learning-based; and (2) deep learning-based.

Morphology-based Skull Stripping methods employ mathematical morphological operations and thresholding techniques for edge detection to extract image features and identify brain surfaces [9]. Deformable surface-based methods evolve and deform a dynamic curve driven by a function of energy towards the active contour. This contour shrinks to take the shape of the brain surface, stripping the skull from the scan. Brain Extraction Tool (BET) by Smith et al. [10] and BET2 by Jenkinson et al. [11] are the most common deformable surface-based methods due to their inclusion in the FMRIB Software Library (FSL) [12]. Intensity-based methods differentiate regions of the brain by focusing on pixel-wise intensities. The main disadvantage of such an approach is that it is sensitive to intensive bias present in the scan, which may be caused due to fluctuations in the magnetic field while scanning [13]. Template and atlas-based methods strip the skull using a template or atlas registration of the brain. It can distinguish between brain tissues even when no pre-defined relationship between the values has been defined [14]. Hybrid methods combine one or more of the abovementioned approaches to improve upon the accuracy of the results [15].

Machine and deep learning methods carry great potential for procedures such as Skull Stripping. Unlike conventional methods, these methods do not require many user-dependent parameters and can be optimized to work with datasets to achieve above-par results. Unsupervised methods such as the Fuzzy Active Surface Model by Kobashi et al. [16] have been experimented with and have yielded good outcomes. ROBEX, a hybrid generative-discriminative algorithm, explores the MRI image and identifies the most likely contour suggested by the discriminative model [17]. Deep learning methods are carried out using Convolutional Neural Networks (CNNs) and are a highly active research field. These methods are usually implemented through two main approaches: (1) voxel-wise network; and (2) segmenting the complete image as one single feed-forward step. The 3D-CNN method proposed by Kleesiek et al. in 2016 is often referred to as one of the bases of the deep learning methods for skull-stripping [18]. Ronneberger et al. [19] introduced U-Net convolutional networks for biomedical image segmentation, which are detailed in this study. Salehi et al. experimented with auto-context analysis algorithms for brain extraction [20]. Chen et al. [21] presented an auto-context VoxResNet approach, a residual network architecture [22] for brain segmentation.

In the extensive literature reviewed, no studies focused on data augmentation techniques to improve the performance of U-Net architectures for Skull Stripping. The motivation of this study is to employ data augmentation transforms to mimic the conditions posed by the different scanning parameters associated with each scan. This research evaluates the performance of Vanilla, Residual, and Dense 2D U-Net architectures for Skull Stripping. Dense 2D U-Net architectures have not previously been employed for this task and hence are emphasized and focused on in the scope of this study.

The research conducted is presented in three sections. The following section describes the methodology of the study, including the overview, network architectures, pre-processing, augmentation, and implementation. The subsequent section elaborates on the results and discusses them as a comparative study, followed by the conclusion.

## 2  Methodology

Skull Stripping involves separating brain tissue, including grey matter and white matter, from non-brain voxels such as the skull, scalp, and dura mater. Many assessments require Skull Stripping in the initial stages of the processing pipeline, such as volumetric or longitudinal analysis for diagnostics [23] [24], analysis of the progression of multiple sclerosis [25], cortical and sub-cortical analysis [26], assessment of mental disorders [27], and for the planning of surgical interventions in neurology [28]. The methodology employed in this study is described in the next section and can be summarized as shown in Figure 2.

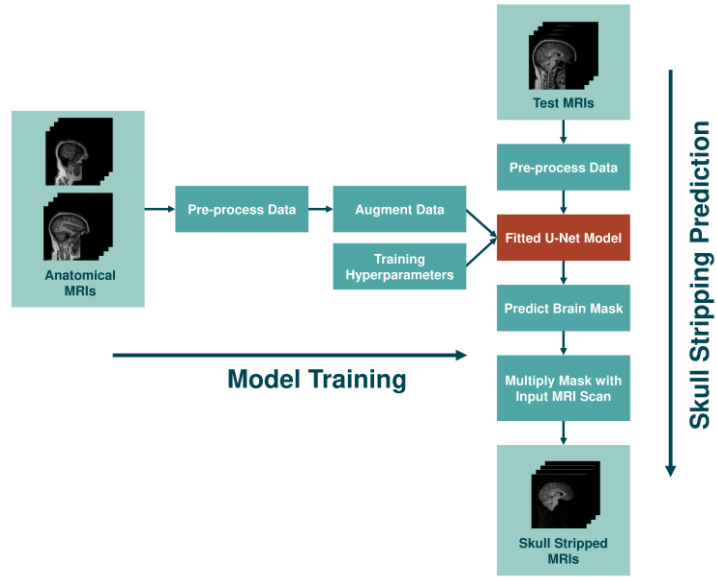

**Fig. 2.** Methodology for Skull Stripping using U-Net Architectures.

### 2.1  Overview

The general methodology employed for this study is divided into multiple steps as a part of two main pipelines. The main pipelines are (1) model training; and (2) Skull Stripping prediction on a test MRI. Both pipelines are initialized by a pre-processing step to z-normalize the data, detailed in Section 2.3. MRI scans are highly susceptible to bias variations due to inhomogeneity in the magnets [29]. Normalizing the data ensures that the mean of the data lies at 0 and that no particular regions or scans are given a higher priority than others. The training data is augmented to increase the samples fivefold. After defining the training hyperparameters for the chosen U-Net architecture, the models are fit to the augmented data, which are then used to predict Skull Stripped masks of unseen MRI head scans. The mask generated is then multiplied with the input MRI to generate a Skull Stripped scan. The following section details the data used for the research study.

## 2.2 Dataset

The Neurofeedback Skull-stripped (NFBS) repository dataset [7] is used to train the three architectures of U-Net in this repository. The repository is available to researchers through the Pre-processed Connectomes Project. The database comprises 125 T1-weighted structural MRI scans and brain masks for each. The 125 MRIs are first segmented using the semi-automated BEaST method [30] and then manually correcting the improper results. In total, 85 of the resultant images are further manually rectified, while the rest are perfectly segmented without human assistance.

Each MRI in the dataset is formed by a sequence of 192 sagittal slices with 256 × 256 mm$^2$ field of view and an acquisition resolution of 1 × 1 × 1 mm$^3$ [31]. This anatomical data for each scan and mask is stored as NIfTI files, a standard imaging format for neuroradiology research. Of these scans, 110 are used for training, and the remaining 15 are used for testing.

## 2.3 Pre-processing and Data Augmentation

Z-normalization of intensity values can increase the mean accuracy of T1-weighted classifiers, as shown in the study by Carré et al. [32]. This method normalizes images by subtracting the mean intensity value of the brain pixels ($\mu_{brain}$), from each pixel intensity ($I(x)$) and dividing the result by the standard deviation of the brain pixels ($\sigma_{brain}$), shown in Equation (1):

$$I_{Z-norm}(x) = \frac{I(x) - \mu_{brain}}{\sigma_{brain}} \quad (1)$$

Data augmentation is a standard procedure employed in deep learning techniques to enhance the size and quality of training datasets such that better models can be built using them. It can help increase the size of the dataset manifold and tune the model to be robust to multi-scanner variations in the data. This study uses a combination of spatial and intensity transforms to expand the training MRI dataset size fivefold, from 110 initial scans to 550 augmented scans of 192 sagittal slices each. This increases the total slices available for training to 105,600 individual sagittal slices.

## 2.4 Neural Network Architectures

In this study, three neural network architectures are evaluated for Skull Stripping of augmented MRI head scans: (1) Vanilla 2D U-Net Architecture; (2) Residual 2D U-Net Architecture; and (3) Dense 2D U-Net Architecture.

**Vanilla 2D U-Net Architecture.** The contractive path of the Vanilla 2D U-Net architecture employs basic 2D convolutions. It consists of two ReLU-activated 3 × 3 convolutional layers with same padding, followed by a 2 × 2 max-pooling layer with stride (2, 2). This block is repeated four times before beginning the expansive path. This path consists of 2 × 2 up-convolutions or transpose convolutions to generate an up-sampled feature map created by the contractive path and a shortcut concatenation with the corresponding feature map from the contracting path. This is followed by two $3 \times 3$ convolutional layers with Rectified Linear Unit (ReLU) activation. At the final layer, a unit convolution with sigmoid activation is used to classify each intensity vector as either brain or non-brain tissue. This architecture has a total of 7,759,521 trainable parameters.

**Residual 2D U-Net Architecture.** Residual connections improve the flow of information in the network. It encourages feature reuse through the layers in the form of learning residual functions to the layer inputs [22]. There are four residual block connections in the Residual 2D U-Net Architecture contractive path. In each of these blocks, the input to the two $3 \times 3$ convolutional layers is concatenated with their output. This concatenated result is then passed through a max-pooling layer with a $2 \times 2$ kernel and a stride of (2, 2). The same residual block pattern is also applied to the expansive path, with the difference that instead of max-pooling layers, up-convolutions or transpose convolutions are used to upscale the feature image. Similar to the Vanilla 2D U-Net, shortcut connections are made across the contractive and expansive paths at the same levels, and the expansive path ends with a unit convolution and a sigmoid activation function for multi-class segmentation. The total number of training parameters for this architecture is 9,895,073.

**Dense 2D U-Net Architecture.** According to the literature reviewed, the Dense 2D U-Net architecture for Skull Stripping proposed in this research study has not been explored previously. The groundwork for building Dense U-Net architectures is laid by Huang et al. in their work on Dense Convolutional Networks [33]. It allows for feature reuse throughout the network and strengthens shallower models with the benefits of deeper networks. They encourage identity mappings, deep supervision, and diversified depth for the models. The dense skip connections allow for improved aggregation of features across different semantic scales [34]. In the Residual 2D U-Net architecture, the input to the two $3 \times 3$ convolutional layers is concatenated with their output. In the Dense 2D U-Net architecture, along with these residual connections, additional dense connections are established that run within these blocks of convolutional layers. The concatenation of the output from the first convolutional layer and the input is fed through another convolutional layer. The total number of trainable parameters in this network is 15,479,681.

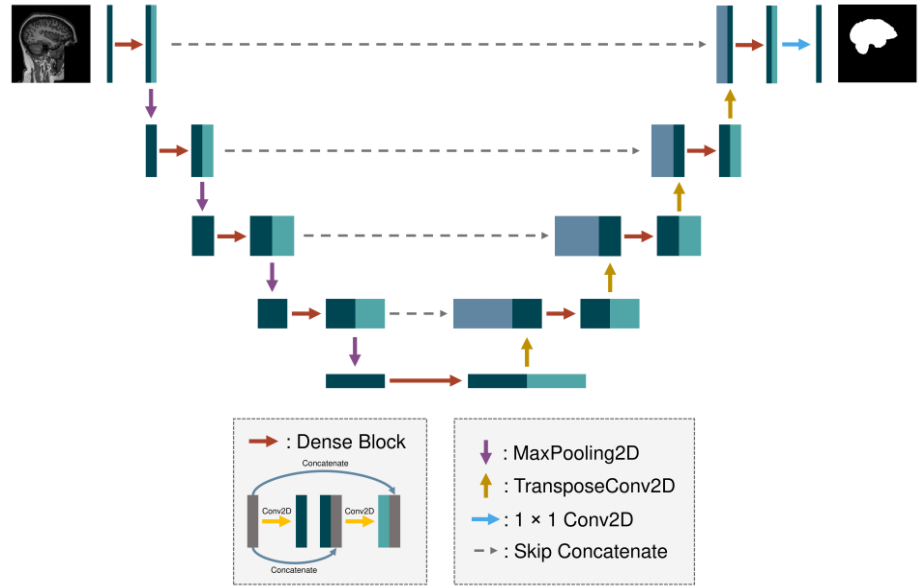

**Fig. 3.** Diagrammatic Representation of the Dense 2D U-Net Architecture for Skull Stripping.

### 2.5 Implementation

Each augmented MRI scan can be realized as an array of shape $192 \times 256 \times 256$. To optimize the training for our available resources, the data is converted from Float64 to Float16. The masks are converted from Float64 to Int8. The processed data is saved in the standard binary NumPy file format (NPY). This format stores the shape and data type information necessary to reconstruct the data on any machine, regardless of the system architecture. The dataset is loaded as a hashed memory map to preserve RAM for training at the expense of computational latency.

The project is implemented with a TensorFlow [35] backend, using the Keras APIs [36] included in the package. All models are trained on Google Colab[1], using an NVIDIA A100 Tensor Core GPU (40GB)[2] for hardware acceleration. The Adam optimizer [37] is employed for network training. The summary of hyperparameters used for training the models is shown in Table 1.

The train/validation split is set to 90/10, with a repeated holdout methodology. The EarlyStopping callback in TensorFlow is used to prevent overfitting of the models to the training dataset. The callback monitors the validation loss of the model and stops training when the validation loss starts increasing over epochs, with a patience value of 2. The batch size for each model is chosen according to memory limitations. In the next section, the training progress results are summarized and discussed.

---

[1] Google Corp., Mountain View, CA USA, 2018
[2] NVIDIA Corp., Santa Clara, CA USA, 2021

**Table 1.** Summary of Training Hyperparameters.

| Hyperparameter | Details |
|---|---|
| Optimizer | Adam |
| $\beta_1$ | 0.9 |
| $\beta_2$ | 0.999 |
| $\varepsilon$ | $10^{-8}$ |
| Learning Rate | $10^{-5}$ |
| Learning Rate Decay | $1.99 \times 10^{-7}$ |

## 3 Results and Discussion

This section presents the performance of the three neural network architectures in the form of loss function and accuracy. Binary cross entropy loss function and an accuracy metric are used to evaluate the performance of the models. The summary of training parameters and training, validation, and testing results are shown in Table 2. The three architectures trained through 9 to 10 epochs before exiting the process due to the EarlyStopping callback. The loss functions of the Residual and Dense 2D U-Net architectures converged in 9 epochs compared to the 10 epochs required by the Vanilla 2D U-Net architecture.

**Table 2.** Summary of Training Process Results.

| Architecture | Epochs | Batch Size | Training Loss | Training Accuracy | Validation Loss | Validation Accuracy | Testing Loss | Testing Accuracy |
|---|---|---|---|---|---|---|---|---|
| Vanilla 2D U-Net | 10 | 32 | 0.0066 | 99.72% | 0.0093 | 99.63% | 0.0065 | 99.73% |
| Residual 2D U-Net | 9 | 32 | 0.0066 | 99.72% | 0.0092 | 99.63% | 0.0067 | 99.72% |
| Dense 2D U-Net | 9 | 16 | 0.0053 | 99.77% | 0.0085 | 99.67% | 0.0062 | 99.75% |

The epoch-wise loss and accuracies for the three neural network architectures are monitored during training and validation and represented graphically in Figure 4 (a), (b), and (c). Upon training completion, the models are tested on the 15 pre-processed test MRIs kept aside initially.

3D U-Net architectures are also common along with their 2D counterparts. Theoretically, 3D convolutions leverage the spatiotemporal information between individual slices to add another correlational element to the model [38]. Research activities by Hwang et al. [39] and Kolařík et al. [40] have used 3D U-Net architectures for Skull Stripping and provided results better than conventional methods. Hsu et al. employed 3D U-Net architectures for Skull Stripping of rat head scans, which performed well in various metrics such as Dice, Jaccard, center-of-mass distance, and Hausdorff distance. The 3D U-Net is inclined to have higher sensitivity but lower positive predictive values, as it misclassifies a higher ratio of nonbrain tissues than brain tissues. Hence, fewer false positive errors in 2D U-Nets lead to higher precision. A 3D U-Net16 architecture also causes more loss of original 2D dimensional information as compared to a 2D U-Net64 architecture. Therefore, 2D U-Nets perform better than 3D U-Nets with a similar number of parameters [41].

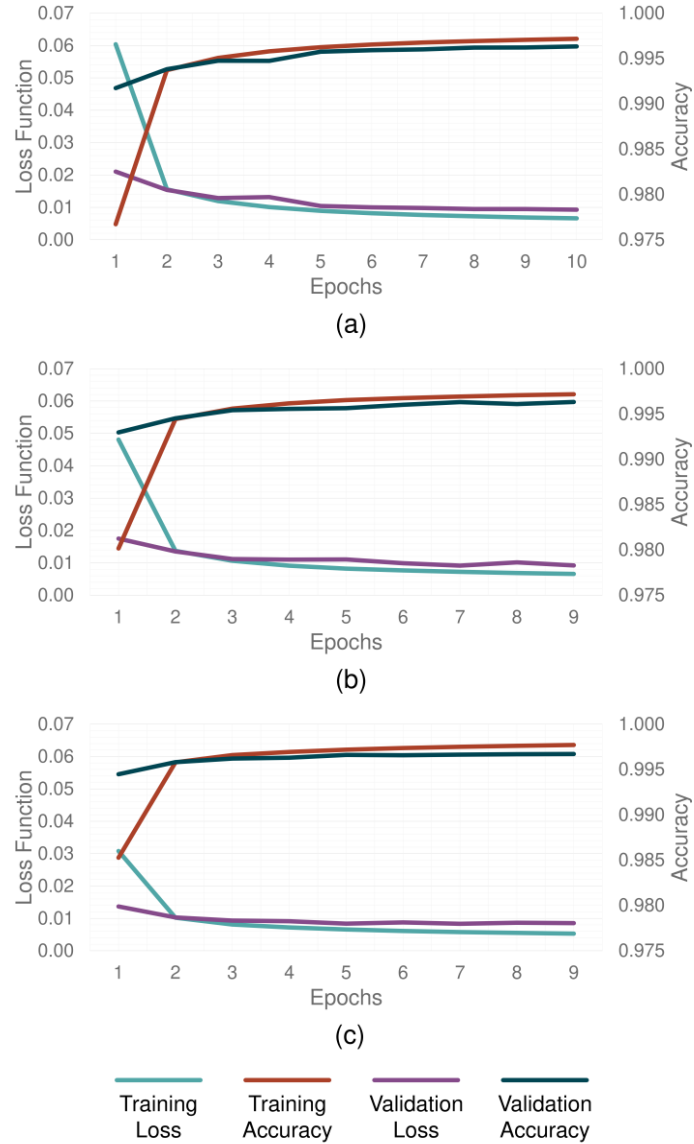

**Fig. 4.** Graphical representation of training and validation accuracy and losses for (a) Vanilla; (b) Residual; and (c) Dense 2D U-Net Architectures.

The performance of Dense 2D U-Net architectures proposed in this research study is promising and can be experimented with further by introducing further augmentation techniques, deepening the network, choosing different evaluation metrics, and tuning the chosen hyperparameters.

## 4   Conclusion

The performances of Vanilla, Residual and Dense 2D U-Net architectures have been evaluated for Skull Stripping of MRI head scans. The accuracy metrics of the three models have been compared on a test dataset. The results of the Dense architecture surpass the rest, the reasons for which have been justified in the study. These algorithms can be employed for many neuroimaging tasks that require Skull Stripping, such as a pipeline to assess the volumetric changes in the brain of patients diagnosed with disorders and diseases such as Major Depressive Disorder (MDD) and Alzheimer's Disease (AD), or to assess the progression of conditions such as Multiple Sclerosis. In the future, the performance of existing and novel 3D U-Net architectures can be compared against that of the ones discussed here.

## 5   Acknowledgments

We thank COEP Technological University's authorities and the Center of Excellence present at their Department of Electronics and Telecommunication for the support provided towards this study.